\documentclass[twocolumn,showpacs,preprintnumbers,amsmath,amssymb]{revtex4}

\usepackage{graphicx}
\usepackage {amsmath}
\usepackage{hyperref}

\begin{document}

\title{Effective Hamiltonian study of excitations in a boson- fermion mixture with attraction
between components}

\author{A.M. Belemuk}
\affiliation{Institute for High Pressure Physics, Russian Academy
of Sciences, Troitsk 142190, Moscow Region, Russia}

\author{V.N. Ryzhov}
\affiliation{Institute for High Pressure Physics, Russian Academy
of Sciences, Troitsk 142190, Moscow Region, Russia}

\date{\today}

\begin{abstract}
An effective Hamiltonian for the Bose subsystem in the mixture of
ultracold atomic clouds of bosons and fermions with mutual
attractive interaction is used for investigating collective
excitation spectrum. The ground state and mode frequencies of the
$^{87}$Rb and $^{40}$K mixture are analyzed quantitatively at zero
temperature. We find analytically solutions of the hydrodynamics
equations in the Thomas- Fermi approximation. We discuss the
relation between the onset of collapse and collective modes
softening and the dependence of collective oscillations on
scattering length and number of boson atoms.
\end{abstract}

\pacs{03.75.Lm,03.75.Kk,67.57.Fg}

\maketitle

\section{Introduction}

Bose-Einstein condensation (BEC) in ultracold trapped atomic gases
\cite{[1],[2],[3]} has been the subject of intense theoretical and
experimental interest \cite{Bloch08, Giorgini08}. Experimental
studies of the BEC properties in confined vapours of alkali atoms
have been extended to double Bose condensates
\cite{cr1,cr2,cr3,cr4}, to the achievement of quantal degeneracy
in gases of fermionic atoms \cite{sci99}, and to dilute mixtures
of Bose and Fermi particles \cite{sci01,prl1,NaLi,Modugno02, KRb}.

Collisional interaction between bosons and fermions greatly
affects the properties and leads to a rich phase diagram of the
degenerate mixtures. Theoretical considerations predict the
phenomenon of component separation for systems with a positive
coupling constant \cite{Nygaard99, Yi01, Minguzzi2000, Akdeniz02,
Capuzzi02}, instabilities and significant modification of the
properties of individual component in a case of boson-fermion
attraction \cite{Roth02, Molmer98, jezek}, the formation of a
superfluid state due to boson-induced fermion-fermion attraction
\cite{Stoof2000}. The simultaneous collapse of the two species has
been observed experimentally in the $^{40}$K - $^{87}$Rb mixture
by Modugno and co-workers \cite{Modugno02} as a sudden
disappearance of fermion cloud when the number of bosons is
increased over an instability value $N_{Bc}\approx 10^{5}$, the
critical number of fermions being $N_K\approx 2\times 10^{4}$. The
stability diagram for the $^{40}$K - $^{87}$Rb mixture and the
critical particle number for the onset of the collapse has been
considered in \cite{Ospelkaus06}.

The dynamical properties have been investigated for various
boson-fermion mixtures, i.e., $^{39}$K -- $^{40}$K
\cite{Capuzzi01, Maruyama05}, $^6$Li -- $^7$Li \cite{Capuzzi03,
Capuzzi03_2}, $^{87}$Rb -- $^{40}$K \cite{Capuzzi04}. The
theoretical studies of the collective oscillations based on a
random-phase approximation \cite{Capuzzi01, Capuzzi03,
Capuzzi03_2}, direct numerical integration of the time-dependent
equations \cite {Maruyama05}, semi-analytical methods
\cite{Miyakawa2000, Sogo02}, or on numerical solutions of the
coupled eigenvalue equations for boson and fermion density
fluctuations \cite{Capuzzi04}.

The analytic treatment of the elementary excitations for Bose-
condensed gases confined in magnetic traps has been given by
Stringari \cite{Stringari96}. The Stringari's dispersion law of
the discretized normal modes does not depend on the interparticle
interaction strength and holds in the hydrodynamic limit $N_B
a_B/a_{ho} \gg 1$, where $a_{ho}= (\hbar/m_B\omega_B)^{1/2}$ is
the harmonic oscillator length, $a_B > 0$ is the bosonic
scattering length, and $N_B$ is the number of bosons. The
numerical confirmation of these behavior has been given in
\cite{Edwards96} by applying linear-response theory. In the case
of attractive interaction ($a_B < 0$) the system becomes more
compressible when approaching the critical number $N_{Bc}$ for
collapse. The $N_{B}$ dependence of the lowest monopole frequency
$\omega_M$ can be determined analytically \cite{Ueda98} using a
variational calculation of the ground state based on Gaussian
trial wave function for the order parameter. It vanishes as
$\omega_M \sim \left(1- N_B/N_{Bc} \right)^{1/4}$.

Collective modes of a boson-fermion mixture with mutual attractive
interaction have been studied by Capuzzi and coworkers
\cite{Capuzzi04} using the equations of generalized hydrodynamics.
The numerical procedure included the decomposition of the density
fluctuations into components of definite angular momentum $l$,
i.e. $\delta \rho({\bf r})= \delta \rho_l(r) Y_{lm}(\theta,
\varphi)$, and solving an eigenvalue problem for coupled equations
in a given $l$ subspace by means of standard linear-algebra
routines. It was found that the collective spectra show a
frequency softening of a family of modes of bosonic nature as a
signature of the incipient collapse. This softening becomes most
pronounced in a very narrow region of parameters near collapse.

The purpose of the present work is to present an analytical method
for studying the dynamics of a harmonically confined Bose- Fermi
clouds and provide a theoretical discussion of interaction effects
on the collective oscillations. We analyze quantitatively
dynamical properties of the $^{87}$Rb and $^{40}$K mixture with an
attractive interaction between bosons and fermions at $T=0$ . The
dynamics of a boson-fermion mixture is found amenable to analytic
approaches by effective Hamiltonian method \cite{ChRyz04,cr5}.
This method well describe the density profiles of the Bose- Fermi
mixtures even close to collapse  and predicts the correct value
for critical particle number to make the collapse diagram
\cite{Ospelkaus06}. In order to gain more physical insight and to
solve analytically the linearized hydrodynamic equations we
consider a special limiting case in which the kinetic energy term
is negligible compared to confining and boson- boson interaction
energy, the Thomas- Fermi (TF) approximation for equilibrium
condensate profile similar to Stringari's approach
\cite{Stringari96}. But in contrast to \cite{Stringari96} our
spectrum explicitly depends on the ratio of the boson- boson and
boson- fermion couplings. We show a signature of the softening of
the bosonic collective mode spectra in the vicinity of a collapse,
resulting from peculiar property of the ground state density when
the system become more and more compressible in the vicinity of
the collapse.

The paper is organized as follows. In Sec. II, we present our
effective Hamiltonian approach and introduce necessary notations.
In Sec. III we give our results for ground state boson density
distribution and collective mode spectrum. We then summarize in
Sec. IV.

\section{Theoretical model}
\subsection{Effective interaction}

Our starting point is the functional-integral representation of
the grand-canonical partition function of the Bose-Fermi mixture.
It has the form \cite{popov1,stoof1}:
\begin{multline}
Z= \int D[\phi^*]D[\phi]D[\psi^*]D[\psi]
\exp\left\{-\frac{1}{\hbar} \left[
S_B(\phi^*,\phi)+\right.\right. \\
+ \left.\left. S_F(\psi^*,\psi)+ S_{int}(\phi^*,\phi,\psi^*,\psi)
\right] \right \},
\end{multline}
and consists of an integration over a complex field
$\phi(\tau,{\bf r})$, which is periodic on the imaginary-time
interval $[0,\hbar\beta]$, and over the Grassmann field
$\psi(\tau,{\bf r})$, which is antiperiodic on this interval. The
field $\phi(\tau,{\bf r})$ describes the Bose component of the
mixture, whereas $\psi(\tau,{\bf r})$ corresponds to the Fermi
component. The term describing the Bose gas has the form:
\begin{multline}
S_B(\phi^*,\phi)= \int_0^{\hbar\beta}d\tau\int d{\bf r} \left\{
\phi^*(\tau,{\bf r}) \left( \hbar \frac{\partial}{\partial
\tau}-\frac{\hbar^2\nabla^2}{2 m_B} + \right.\right. \\
+ \left.\left. V_B({\bf r})- \mu_B \right) \phi(\tau,{\bf r})+
\frac{g_B}{2}|\phi(\tau,{\bf r})|^4 \right\}.
\end{multline}
Because $s$- wave collisions between fermionic atoms in the same
hyperfine state are forbidden by the Pauli principle the Fermi-gas
term can be written in the form:
\begin{multline}
S_F(\psi^*,\psi)= \int_0^{\hbar\beta}d\tau\int d{\bf
r}\left\{\psi^*(\tau,{\bf r})\left(\hbar \frac{\partial}{\partial
\tau}-\frac{\hbar^2\nabla^2}{2 m_F}+ \right.\right. \\
+ \left.\left.V_F({\bf r}) -\mu_F\right)\psi(\tau,{\bf r})
\right\}.
\end{multline}
Here $V_{B,F}({\bf r})$ are the external confining potentials, and
$\mu_{B,F}$ are the chemical potentials for the Bose- and Fermi-
components respectively. Under isotropic harmonic confinement
$V_{B,F}({\bf r})= m_{B,F}\omega_{B,F}^2 r^2/2$, $m_B$ and $m_F$
are the masses of bosonic and fermionic atoms respectively. The
trap parameters $\omega_B$ and $\omega_F$ are chosen in such a way
that $m_B\omega_B^2/2= m_F\omega_F^2/2= V_0$, that is why
$\omega_F= \sqrt{m_B/m_F} \omega_B$. The radius of the Fermi-
cloud $R_F$ can be estimated as $V_0 R_F^2 \simeq \mu_F$.

The term describing the interaction between the two components of
the Fermi- Bose mixture is:
\begin{equation}
S_{int}(\phi^*,\phi,\psi^*,\psi)=g_{BF}\int_0^{\hbar\beta}d\tau\int
d{\bf r} |\psi(\tau,{\bf r})|^2|\phi(\tau,{\bf r})|^2,
\end{equation}
where $g_B=4\pi \hbar^2a_B/m_B$ and $g_{BF}=2\pi
\hbar^2a_{BF}/m_I$, $m_I=m_B m_F/(m_B+m_F)$, $a_B$ and $a_{BF}$
are the $s$- wave scattering lengths of boson-boson and
boson-fermion interactions.

The integral over Fermi fields
\begin{equation}
Z_F= \int D[\psi^*]D[\psi] e^{-\frac{1}{\hbar}\left
(S_F(\psi^*,\psi)+ S_{int}(\phi^*,\phi,\psi^*,\psi)\right)}
\end{equation}
is Gaussian, we can calculate this integral and obtain the
partition function of the Fermi system as a functional of Bose
field $\phi(\tau, {\bf r})$ which for $T= 0$ has the form
\cite{ChRyz04,cr5}:
\begin{gather}
Z_F= \exp\left(-\frac{1}{\hbar}S_{eff}\right), \quad S_{eff}= \int
\limits_0^{\hbar\beta}d\tau \int d{\bf r} f_{eff}(|\phi(\tau,{\bf
r})|), \notag\\
f_{eff}(|\phi(\tau,{\bf r})|)= -\frac25\varkappa \left[ \mu_F-
V_F({\bf r})- g_{BF}|\phi(\tau,{\bf r})|^2 \right]^{5/2},
\end{gather}
where $\varkappa= \sqrt{2}m_F^{3/2}/(3\pi^2\hbar^3)$.

Using the fact that due to the Pauli principle (quantum pressure)
the radius of the Bose condensate is much less than the radius of
the Fermi cloud $R_F\approx \sqrt{\mu_F/V_0}$, one can use an
expansions in powers of $V_F({\bf r})/\mu_F$ and obtain the
effective Hamiltonian in the form \cite{ChRyz04,cr5}:
\begin{gather}
Z= \int D[\phi^*]D[\phi] {\:} e^{-\frac1{\hbar}(S_B+ S_{eff})},
\quad S_B+ S_{eff}= \nonumber\\
= \int \limits_0^{\hbar \beta} d\tau \left[ \int d^3 r {\,}
\phi^*(\tau, {\bf r}) \hbar \frac{\partial}{\partial \tau}
\phi(\tau, {\bf r})+ H_{eff}[\phi^*, \phi] \right],
\end{gather}
\begin{multline} \label{he}
H_{eff}[\phi^*, \phi]= \int d{\bf
r}\left\{\frac{\hbar^2}{2m_B}|\nabla\phi|^2+ \right.
\\
+ (V_{eff}({\bf r})-\mu_B)|\phi|^2 +
\left.\frac{g^{BB}_{eff}}{2}|\phi|^4+
\frac{g_{eff}^{BF}}{3}|\phi|^6\right\},
\end{multline}
where
\begin{gather}
V_{eff}({\bf r})= k_0 \frac{m_B \omega_B^2}{2} r^2, {\;} k_0= (1-
\frac{3}{2}\varkappa \mu_F^{1/2} g_{BF}), \nonumber \\
g^{BB}_{eff}= g_B- \frac{3}{2}\varkappa \mu_F^{1/2} g_{BF}^2,
\quad g_{eff}^{BF}= \frac{3\varkappa}{8\mu_F^{1/2}}g_{BF}^3.
\end{gather}
The first three terms in (\ref{he}) have the conventional
Gross-Pitaevskii \cite{Dalfovo99} form, and the last term is a
result of boson-fermion interaction. It corresponds to the
three-particle \emph{elastic} collisions induced by the
boson-fermion interaction. In contrast with \emph{inelastic}
3-body collisions which result in the recombination and removing
particles from the system \cite{Kagan9698}, this term for
$g_{BF}<0$ leads to increase of the gas density in the center of
the trap in order to lower the total energy.

\subsection{Hydrodynamic approach}

Now rewrite the action in terms of hydrodynamic variables density
and phase $\phi(\tau, {\bf r})= \sqrt{\rho(\tau, {\bf
r})}e^{i\theta(\tau, {\bf r})}$
\begin{multline} \label{Zeff1}
Z= \int D[\rho] D[\theta] {\:} e^{-\frac{1}{\hbar}(S_B+
S_{eff})}, \\
S_B+ S_{eff}= \int\limits_0^{\hbar \beta} d\tau \int d^3r \left[
i\hbar \rho \frac{\partial \theta}{\partial \tau}+ \frac{\hbar^2
\rho}{2m_B} (\nabla \theta)^2+ \right.\\
+ \left. \frac{\hbar^2}{8m_B} \frac{(\nabla \rho)^2}{\rho}+
(V_{eff}({\bf r})- \mu_B)\rho+ \frac{g^{BB}_{eff}}{2}\rho^2+
\frac{g_{eff}^{BF}}{3}\rho^3 \right].
\end{multline}

To simplify the formalism we introduce dimensionless variables
rescaled by the natural quantum harmonic oscillator units of
length $a_{ho}= \sqrt{\hbar/m_B\omega_B}$, and energy
$\hbar\omega_B$: ${\bf r}= a_{ho} {\bf r'}$, $\tau=
\tau'/\omega_B$, $E= \hbar \omega_B E'$, $\rho= \rho'/a^3_{ho}$,
$P= \hbar \omega_B P'/a^3_{ho}$, $S= S'/\hbar$. Then the effective
Hamiltonian takes the form (the primes omitted)
\begin{multline} \label{E}
S_B+ S_{eff}= \int\limits_0^{\beta} d\tau \left[ \int d^3r {\:} i
\rho \frac{\partial \theta}{\partial
\tau}+ H_{eff}[\rho, \theta]  \right], \\
H_{eff}[\rho, \theta]= \int d^3r \left[  \frac{\rho}{2} (\nabla \theta)^2+
\frac{1}{8} \frac{(\nabla \rho)^2}{\rho}+ \right.\\
+ \left. (V_{eff}({\bf r})- \mu_B)\rho+ \frac{u}{4}\rho^2+
\frac{v}{6}\rho^3 \right],
\end{multline}
where $V_{eff}({\bf r})= k_0{\,} r^2/2$, and we introduced
dimensionless parameters for boson- boson and boson- fermion
couplings $u= 2 g^{BB}_{eff}/a_{ho}^3 \hbar\omega_B$ and $v=
2g_{eff}^{BF}/a_{ho}^6 \hbar\omega_B$.

To obtain equation of motion of Bose condensate density and phase
one varies the action $S_B+ S_{eff}$ and goes to the real time
$\tau= it$ \cite{stoof1}. Variation over the phase $\theta(\tau,
{\bf r})$
\begin{equation}
\frac{\delta (S_B+ S_{eff})}{\delta \theta}= 0, \qquad
\frac{\partial \rho}{\partial t}- \frac{\delta H_{eff}}{\delta
\theta}= 0,
\end{equation}
yields the continuity equation
\begin{equation} \label{equ1}
\frac{\partial \rho}{\partial t}+ {\rm div}(\rho {\bf v})= 0,
\end{equation}
where ${\bf v}= \hbar/m [\nabla \theta({\bf r}, t)]$. Variation
over the density $\rho(\tau, {\bf r})$
\begin{equation}
\frac{\delta (S_B+ S_{eff})}{\delta \rho}= 0, \qquad
\frac{\partial \theta}{\partial t}+ \frac{\delta H_{eff}}{\delta
\rho}= 0,
\end{equation}
yields the equation of motion for the phase
\begin{gather}
\frac{\partial \theta}{\partial t}+ \frac{{\bf v}^2}{2}+
\widetilde \mu - \mu_B= 0, \label{ph_mot} \\
\widetilde \mu= - \frac{\nabla^2 \sqrt{\rho}}{2\sqrt{\rho}}+ k_0
\frac{r^2}{2}+ \frac{u}{2} \rho+ \frac{v}{2} \rho^2.
\label{mu_til}
\end{gather}
Taking the gradient we obtain quantum Bernoulli's equation
\cite{Fetter96, Dalfovo99}
\begin{equation}  \label{equ2}
\frac{\partial {\bf v}}{\partial t}+ \nabla \left( \frac{{\bf
v}^2}{2}+ \widetilde \mu \right)= 0. \\
\end{equation}
Equations \eqref{equ1}, \eqref{equ2} are the modified hydrodynamic
equations of Bose- fluid, which incorporates the effect of Bbose-
Fermi interaction. They correspond to the equation of state of the
mixture where pressure and density are related by
\begin{equation}
P= \frac{1}{4} u \rho^2+ \frac{1}{3} v \rho^3+ P_{kin}(\rho),
\end{equation}
where $P_{kin}(\rho)$ is a kinetic energy contribution associated
with spatial variations of the condensate density $\rho({\bf r})$.

Now to obtain elementary excitations one linearizes Eqs.
\eqref{equ1}, \eqref{equ2} around ground state solution
$\rho_0({\bf r})$, ${\bf v}_0= 0$ and looks for eigenvalue mode
$\delta \rho({\bf r},t)= \delta \rho({\bf r}) e^{-i\omega t}$. The
hydrodynamic amplitudes can be combined into a single second-
order equation for the density perturbation \cite{Stringari96}
\begin{equation} \label{eig_inh}
\omega^2 \delta \rho+ {\rm div}(\rho_0\nabla \delta \widetilde
\mu)= 0.
\end{equation}
Solutions of \eqref{eig_inh} give the low- frequency condensate
collective modes of an inhomogeneous Bose component with a local
condensate density $\rho_0({\bf r})$. The Eq. \eqref{eig_inh}
replaces two coupled eigenvalue equations for the density
fluctuations $\delta \rho_{B,F}$ of each species \cite{Capuzzi03,
Capuzzi04}. The coupled eigenvalue equations predicts two sets of
eigenvectors \cite{Capuzzi03, Capuzzi04}, which can be labeled as
fermionic and bosonic ones, according to the nature of their
eigenvalue in the limit of vanishing $g_{BF}$. Solutions of
\eqref{eig_inh} describe oscillations of the bosonic cloud on the
background of the fermionic component and corresponds to that
branch of excitations which has the bosonic origin.

\section{Results and discussion}

\subsection{Ground state}

To clarify the main features of excitation spectrum of a Bose-
Fermi mixture we illustrate our results on the example of
$^{87}$Rb -- $^{40}$K mixture and consider an isotropic trap when
the problem can be treated effectively as a one-dimensional. In
the experiment of Modugno {\it et al.} \cite{Modugno02} K and Rb
atoms experience potentials with an elongated symmetry with
substantial value of trap asymmetry parameter $\lambda=
\omega_z/\omega_{\perp}$. To compare our results with those in the
experiment we have to rescale number of bosons $N_B$ by the
reverse trap asymmetry ratio $N_B \to N_B/\lambda $. For the
chemical potential of an ideal Fermi gas in a trap one can use the
relation $\mu_F= \hbar\omega_F(6\lambda N_F)^{1/3}$
\cite{Butts97}.

The parameters of the $^{87}$Rb and $^{40}$K mixture are the
following \cite{Modugno02}: $N_F= 2\cdot 10^4$, $a_B=5.25$ nm,
$a_{BF}=-21.7^{+4.3}_{-4.8}$ nm. The magnetic potential had an
elongated symmetry, with harmonic oscillation frequencies for Rb
atoms $\omega_{\perp}=\omega_B=2\pi\times 215$ Hz and
$\omega_{B,z}=\lambda\omega_B=2\pi\times 16.3$ Hz. At this
parameter values the reverse trap asymmetry ratio $1/\lambda=
13.2$, characteristic length $a_{ho}= 735$ nm, the chemical
potential for fermions $\mu_F \approx 31{\,} \hbar \omega_B$,
$\omega_F \approx 1.47 {\,}\omega_B$, $k_0= 1.07$, $u= 0.11$, $v=
-0.0003$.

\begin{figure}
\includegraphics[width=8.8cm]{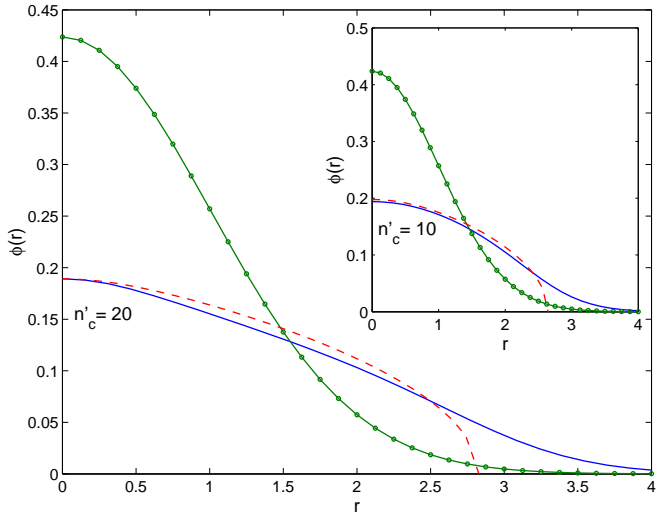}
\caption{\label{Fig1} The profile of the ground state condensate
wave function at $n'_c= 20$ and $n'_c= 10$ (solid line). The line
with circle markers corresponds to the ideal Bose gas in an
external harmonic confinement. Dashed line is for the density
profile in the TF approximation \eqref{TFpr}.}
\end{figure}

The ground state density distribution $\rho_0({\bf r})$ is defined
by the stationary equation ${\bf v}= 0$, $\widetilde \mu = \mu_B$
and gives rise to the modified Gross- Pitaevskii equation
\begin{equation} \label{GP}
- \frac{\nabla^2 \sqrt{\rho_0}}{2\sqrt{\rho_0}}+ k_0
\frac{r^2}{2}+ \frac{u}{2} \rho_0+ \frac{v}{2} \rho_0^2= \mu_B.
\end{equation}

The condensate density $\rho({\bf r}, t)$ is normalized to the
number of atoms in the condensate $\int d^3 r \rho({\bf r}, t)=
N$. In the $T \to 0$ limit, $N$ coincides with the total number of
bosonic atoms in the trap.

Ground state properties of the mixture and numerical solutions of
Eq. \eqref{GP} have been considered in \cite{Ryzhov07,cr6}. It was
shown that the TF approximation gives the good description of
density distribution and BF mixture can be accurately considered
in the TF approximation \cite{PRA06}. In TF approximation
\begin{equation} \label{TFeq}
k_0 r^2- 2\mu_B+ u\rho_0+ v\rho_0^2= 0.
\end{equation}
In this case the density profile has the form
\begin{equation} \label{TFpr}
\rho_0({\bf r})= \frac{n_{cr}}{R_{cr}} \left[R_{cr}- \sqrt{a^2+
k_0 r^2} \right], \quad r \leqslant R_B,
\end{equation}
where $n_{cr}= u/(2|v|)$, $R_{cr}^2= u^2/(4|v|)$, $R_B=
R/\sqrt{k_0}$, $R^2= 2\mu_B$, $a^2= R^2_{cr}- R^2$. In the
limiting case of noninteracting Bose and Fermi clouds ($g_{BF} \to
0$, $v \to 0$, $a \gg 1$, $R_{cr} \approx a+ R^2/2a$, $k_0 \approx
1$, $n_{cr}/2R_{cr}= 1/u$) we recover the TF distribution for
single bose- condensate $\rho_0(r)= (R^2- r^2)/u$.

The parameter $R$ up to a multiplicative factor $1/\sqrt{k_0}$ is
the radius $R_B$ of the Bose condensate. In TF approximation Eq.
\eqref{TFeq} for the center of the trap enables to relate the
value of $R$, i.e. $\mu$, with central density as $R^2= u\rho_0+
v\rho_0^2$. Let us express $R$ through parameter $n'_c$ as
\begin{equation} \label{Rsq}
R^2= n'_c(1- \frac{|v|}{u^2}n'_c).
\end{equation}
The values of the parameter $R$ gains the maximal value $R=
R_{cr}$ at $n'_c= u^2/2|v| \approx 20$.

The numerical solution of Eq. \eqref{GP} for the condensate wave
function $\phi({\bf r})= \sqrt{\rho_0({\bf r})}$ and TF profile
\eqref{TFpr} are presented in Fig.~ \ref{Fig1}. They parameterized
by rescaled central density $n'_c= u n_c$, $n_c= |\phi(0)|^2$.
Note that the wave functions in the figure are normalized to
unity.

Figure \ref{Fig1} shows two characteristic profile of the
condensate wave function with increasing the central density from
$n'_c= 10$ to $n'_c= 20$. For $n'_c \lesssim 20$, the behavior is
characteristic of the Bose gas with repulsion, i.e. the evolution
of the profile corresponds to a monotonic expansion of the boson
cloud with increasing number of bosons. The cloud density becomes
more flat at the trap center, approaching TF analytical solution.
For $n'_c \gtrsim 20$, the solution changes qualitatively: the
central density begins to increase.

\begin{figure}
\includegraphics[width=8.8cm]{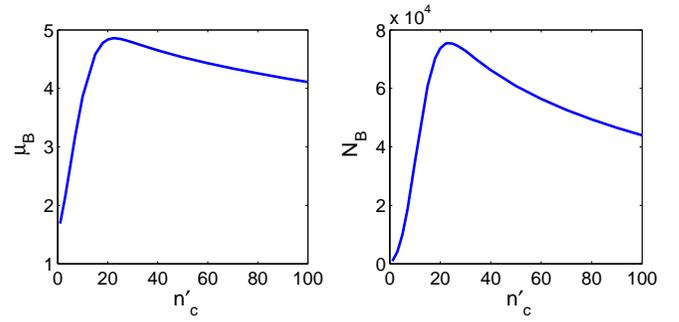}
\caption{\label{Fig2} Chemical potential $\mu_B$ (in the unit of
$\hbar \omega_B$) and number of bosons $N_B$ as functions of the
rescaled central density $n'_c$.}
\end{figure}

For $n'_c \gtrsim 20$, the central density increases significantly
in a small region near the trap center. We attribute the solution
with $n'_c \gtrsim 20$ to the nonstationary states of the
condensate through which the collapse of the condensate wave
function occurs. The value $n'_c \approx 23$ corresponds the
critical particle number $N_{Bc} \simeq 8 \cdot 10^4$ above which
collapse occurs. In that point the branch of stable solutions
meets the branch of unstable solutions of the GP equation. This
qualitative  behavior is the generic signature of a Hamiltonian
saddle node bifurcation \cite{Brachet99}.

Fig.~ \ref{Fig2} shows the boson chemical potential $\mu_B$ and
the number of bosons $N_B$ as function of the parameter $n'_c$,
obtained at numerical solution of Eq. \eqref{GP}. Both curve have
maximal value at $n'_c \simeq 23$. The condition $\partial
N/\partial n'_c= 0$ is a sufficient condition of the zero
excitation mode \cite{Stoof96}. It means that it does not cost
energy to deform the initial density profile continuously into the
final, which indicates the threshold for instability
\begin{multline} \label{def_dens}
N(n'_c+ \delta n'_c)= \int {\:} d^3r {\,} \rho_0({\bf r}, n'_c+
\delta n'_c) \simeq \\
\simeq \int {\:} d^3r {\,} \rho_0({\bf r}, n'_c)= N(n'_c).
\end{multline}

It is interesting to note that the radius of the condensate $R_B=
R/\sqrt{k_0}$ for TF function \eqref{Rsq} qualitatively conform
these behavior, i.e. it increases for $n'_c < n'_{c0}$ and
decreases for $n'_c > n'_{c0}$. There is a critical point where
the dependence $R_B(\mu_B)$ shows the maximum. This remarkable
feature indicates the onset of a collapse in the system
\cite{Ryzhov07}. So we can anticipate some features in the
dynamics of the bose- cloud near the collapse and in the
eigenfrequency spectrum.

\subsection{Collective excitation spectrum}

We now give a brief overview of the calculation of the low-
frequency collective mode spectrum. We have to give explicit
solution of Eq. \eqref{eig_inh}. The main difficulty associated
with Eq. \eqref{eig_inh} is the proper treatment of the
differential operator term ${\rm div} (\rho_0\nabla \delta
\widetilde \mu)$ with
\begin{equation} \label{dmju}
\delta \widetilde \mu= -\frac12  \delta \left(
\frac{1}{\sqrt{\rho}} {\:}\Delta \sqrt{\rho} \right)+ \frac{u}{2}
\delta \rho+ v\rho \delta\rho_0.
\end{equation}
For uniform Bose fluid with constant condensate density $\rho_0$
the first term accounting for quantum pressure reduces to
$-1/(4\rho_0) \Delta \delta \rho$ and the differential operator
term is handled in the straightforward manner
\begin{equation} \label{dif_op_hom1}
{\rm div}(\rho_0\nabla \delta \widetilde \mu)= \rho_0 {\:} \Delta
\delta \widetilde \mu= -\frac{1}{4} {\,} \Delta^2 \delta \rho+
\frac{u \rho_0}{2} \Delta \delta \rho.
\end{equation}
Eigenvalue modes are the plane waves $\delta \rho_{\bf q}({\bf
r})= C e^{i{\bf q r}}$ and the resulting eigenvalues are given by
\cite{Lifshitz80} $\omega_{\bf q}^2= u \rho_0{\bf q}^2/2+ {\bf
q}^4/4$.

For an inhomogeneous Bose- condensate in a trap the TF
approximation must be called on to describe analytically the
condensate collective modes \cite{Stringari96}. For the ground
state the TF density has the form
\begin{equation} \label{TFsingl}
\rho_0(r)= \frac{R^2- r^2}{u}, \qquad r \leqslant R,
\end{equation}
where the radius of Bose- condensate is $R= 2\mu_B$. In Eq.
\eqref{dmju} one omits the quantum pressure term, $\delta
\widetilde \mu \simeq u \delta \rho/2$, and
\begin{equation} \label{dif_op_inhom1}
{\rm div}(\rho_0\nabla \delta \widetilde \mu) \simeq \frac{R^2-
r^2}{2} \Delta \delta\rho+ \nabla \left(\frac{R^2- r^2}{2} \right)
\cdot \nabla \delta \rho
\end{equation}
For a spherical trap, an eigenfunction of Eq. \eqref{eig_inh} can
be written as a product $\delta \rho({\bf r})= \delta \rho_{nl}(r)
Y_{lm}(\theta, \varphi)$ of the radial eigenfunction $\delta
\rho_{nl}(r)$ and a spherical harmonic $Y_{lm}(\theta, \varphi)$,
where $n$ is the radial quantum number. The associated
eigenfunctions are polynomials of the form
\begin{equation} \label{eigStr1}
\delta \rho_{nl}(r)=  C r^l w_{nl}(r^2/R^2) {\:} \Theta(R- r),
\end{equation}
where $\Theta(R- r)$ is the step function. The polynomials
$w_{nl}(y)$ are polynomials of order $n$ and satisfy equation
\begin{equation} \label{hyper_Str}
y(1-y) w''+ \left(\frac{2l+ 3}{2}- \frac{2l+ 5}{2} y \right)w'-
\frac{\epsilon- l}{2} w= 0,
\end{equation}
where $\epsilon= \omega^2$. They can be expressed through the
hypergeometric function $w_{nl}(y)= F(\alpha, \beta, \gamma, y)$
with parameters $\alpha= n$, $\beta= l+ 3/2+ n$, $\gamma= l+3/2$.
The associated energy eigenvalues are found to be
\begin{equation} \label{eigStr2}
\omega_{nl}^2= \omega_B^2(2n^2+ l+ n(2l+ 3)).
\end{equation}
The hypergeometric function $F(\alpha, \beta, \gamma, y)$
satisfies the equation \cite{Abramowitz}
\begin{equation} \label{hyper}
y(1-y) F''+ [\gamma- (\alpha+ \beta+ 1)y]F'- \alpha \beta F= 0.
\end{equation}
The eigenfunctions \eqref{eigStr1} for the Bose-condensed cloud
fluctuations vanish outside the cloud radius $R$ and present a
discontinuity at $R$. The discontinuity is physically acceptable
in view of the fact that the kinetic energy term has been set as
negligible in taking the strong- coupling limit.

Let us now address the issue in the case Bose- Fermi mixture. As
well as in the previous case we omit the quantum pressure term in
\eqref{dmju} and the perturbation for $\widetilde \mu$ takes the
form
\begin{equation}
\delta \widetilde \mu= \frac{u}{2}\delta \rho+ v \rho_0\delta
\rho, \quad \nabla \delta \widetilde \mu= \left(\frac{u}{2}+
v\rho_0 \right) \nabla \delta \rho+ v (\nabla \rho_0) \delta \rho,
\end{equation}
and
\begin{multline} \label{expr_div}
{\rm div}(\rho_0\nabla \delta \widetilde \mu)=
\left(\frac{u}{2}\rho_0+ v\rho_0^2 \right) \Delta \delta \rho+ \\
+ \left(\frac{u}{2} \nabla \rho_0+ \frac32 v \nabla \rho_0^2
\right)\cdot \nabla \delta \rho+ v\rho_0^2 \left[\left(
\frac{\nabla \rho_0}{\rho_0} \right)^2+ \frac{\Delta
\rho_0}{\rho_0} \right] \delta \rho.
\end{multline}
Due to $\nabla \rho_0/\rho_0 \ll 1$ we should omit the last two
terms in square brackets in TF approximation. So we can rewrite
the expression \eqref{expr_div} in the form
\begin{multline} \label{div_op1}
{\rm div}(\rho_0\nabla \delta \widetilde \mu) \simeq \left(
\frac{1}{2}(u\rho_0+ v\rho_0^2)+ \frac{v}{2}\rho_0^2\right) \Delta
\delta \rho+ \\
+ \nabla  \left( \frac{1}{2}\left(u\rho_0+ v\rho_0^2 \right)+
v\rho_0^2\right) \cdot \nabla \delta \rho.
\end{multline}
Now we use in Eq. \eqref{div_op1} the ground state density profile
$\rho_0(\bf r)$ from Eq. \eqref{TFpr}, which simply states that
due to Eq. \eqref{TFeq} $u\rho_0+ v\rho_0^2= 2\mu_B- k_0r^2$ and
come to the result
\begin{multline} \label{div_op2}
{\rm div}(\rho_0\nabla \delta \widetilde \mu) \simeq \left(
\frac{R^2- k_0r^2}{2}+ \frac{v\rho_0^2}{2}\right) \Delta
\delta \rho+ \\
+ \nabla  \left( \frac{R^2- k_0r^2}{2}+ v\rho_0^2 \right) \cdot
\nabla \delta \rho.
\end{multline}
Note that in the $v \to 0$ limit the Eq. \eqref{div_op2}
transforms into Eq. \eqref{dif_op_inhom1} for single Bose
condensate. In principal, the solution of Eq. \eqref{eig_inh} with
approximate ${\rm div} (\rho_0\nabla \delta \widetilde \mu)$
giving by Eq. \eqref{div_op2} should give rise to collective
excitations spectra in TF approximation. But the term $v\rho_0^2$
is the complicated function with square root singularity. In order
give analytical treatment of collective excitations we have to
approximate the $v\rho_0^2$ term by more analytically simple
function. It is reasonable to simplify the $v\rho_0^2$ term and
approximate it by quadratic function on the radius $r$. For
$\rho_0^2$ we use Eq. \eqref{TFpr} and expand the radical near the
center of the trap which gives rise to
\begin{equation} \label{TF3}
\rho_0^2(r) \simeq \frac{1}{|v|} \frac{k_0 b}{a} \left((R'_B)^2-
r^2\right),
\end{equation}
where $R'_B= ab/k_0$ and $b= R_{cr}- a$. In the framework of the
density profile \eqref{TF3} it is possible to give explicitly
solutions of Eq. \eqref{eig_inh} through the hypergeometric
functions.

It is clear that the result \eqref{TF3} does not describe density
near $r= R'_B$. This fact is an artifact of the TF approximation,
arising from the vanishing of $\rho_0({\bf r})$ at $r= R'_B$. The
dependence $\rho_0^2$ in Eq. \eqref{TF3} has slightly different
radius of the cloud $R'_B$ in comparing with $R_B$ in Eq.
\eqref{TFpr}, but this approximation is consistent with the TF
approximation because in the region near boundary density of the
cloud is small. Note that $(R'_B)^2= R^2/k_0-(R^2_{cr}-
\sqrt{R_{cr}^2- R^2})/k_0$ and $R'_B \simeq R_B$ at $R \ll
R_{cr}$. In this regime the density profiles giving by Eqs.
\eqref{TFpr} and \eqref{TF3} are the similar. In contrast in the
critical regime at $R \lesssim R_{cr}$ the parameter $R'_B \ll
R_B$ and the approximation Eq. \eqref{TF3} is no longer valid.

Now for $v\rho_0^2$ term in Eq. \eqref{div_op2} we use expression
\eqref{TF3} and introduce notations $\alpha_1= k_0(1- b/a)$,
$\alpha_2= k_0(1- 2 b/a)$. Then
\begin{equation}
{\rm div}(\rho_0 \nabla \widetilde \mu)= \frac{k_0 R_B^2- \alpha_1
r^2}{2} \Delta \delta \rho- \alpha_2 r \frac{\partial}{\partial r}
\delta \rho.
\end{equation}
Now the equation \eqref{eig_inh} takes the form
\begin{equation} \label{rho_eq}
\frac{R_*^2- r^2}{2} \Delta \delta \rho- \alpha_3 r
\frac{\partial}{\partial r} \delta \rho+ \epsilon \delta \rho= 0.
\end{equation}
where $R_*^2= k_0 R_B^2/\alpha_1$, $\alpha_3= \alpha_2/\alpha_1$,
$\epsilon= \omega^2/\alpha_1$. Separating variables $\delta
\rho({\bf r})= r^l{\,}G(r)Y_{lm}(\theta, \varphi)$ and
substituting $y= r^2/R_*^2$, one obtains for the function $w(y)=
G(r(y))$ the following equation:
\begin{equation} \label{w_eq}
y(1-y)w''+ \left( \frac{2l+3}{2}- \frac{2l+3+ 2\alpha_3}{2}y
\right)w'+ \frac{\epsilon- \alpha_3 l}{2}w= 0.
\end{equation}
If we compare it with analogous Eq. \eqref{hyper_Str} we can see
that it contains additional parameters $\alpha_1$ and $\alpha_2$
including the effect of interparticle interaction. At $v \to 0$
the coefficients $\alpha_1, \alpha_2 \to 1$, and the Eq.
\eqref{w_eq} transforms into Eq. \eqref{hyper_Str} for a pure
boson system. Eq. \eqref{w_eq} is a standard equation of the form
\eqref{hyper} for hypergeometrical function $F(\alpha, \beta,
\gamma, y)$. It should be a polynomial of $n$- th degree, which
for $v \to 0$ tends to Stringari's solution \eqref{eigStr1}. So
the parameters of the hypergeometrical function $F(\alpha, \beta,
\gamma, y)$ get the values $\alpha= -n$, $\beta= (2l+ 1+
2\alpha_3)/2+ n$, and $\gamma= (2l+3)/2$.

The associated energy eigenvalue are found to be (in dimensional
units)
\begin{equation} \label{eigTF}
\omega_{nl}^2= \omega_B^2 \left[\alpha_1(2n^2+ n(2l+ 1))+ \alpha_2
(2n+ l) \right].
\end{equation}
The result \eqref{eigTF} can be considered as a qualitative
signature of shifts of eigenfrequencies in the presence of the
boson- fermion coupling. At $v \to 0$ frequencies \eqref{eigTF}
transform into the Stringari's spectrum \eqref{eigStr2}.

The dispersion law of the normal modes given by the formula
\eqref{eigTF} has the additional coefficients $\alpha_1$ and
$\alpha_2$ which incorporate the effect of interparticle
interaction. In explicit form the coefficients $\alpha_1$ and
$\alpha_2$ are related with $N_B$, $N_F$, and $g_{BF}$ through the
parameters $R$ and $R_{cr}$
\begin{gather} \label{alpha1_2}
\alpha_1= k_0 \left(1- \frac{R_{cr}- \sqrt{R_{cr}^2-
R^2}}{\sqrt{R_{cr}^2- R^2}} \right) \notag \\
\alpha_2= k_0 \left(1- 2\frac{R_{cr}- \sqrt{R_{cr}^2-
R^2}}{\sqrt{R_{cr}^2- R^2}} \right)
\end{gather}
Formally these relations are valid only at $R \ll R_{cr}$. In our
approach we interpolate they up to $R \lesssim R_{cr}$. These
rather crude approximation nevertheless gives rise to the
qualitatively correct picture of collective excitation spectra
tendency to become softening near the collapse transition.

\begin{figure}
\includegraphics[width=8.8cm]{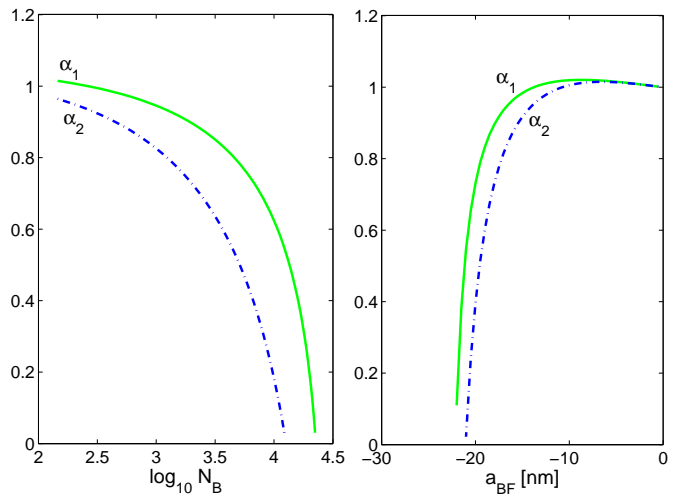}
\caption{\label{Fig3} The coefficients $\alpha_1$ and $\alpha_2$
of the formula \eqref{eigTF} as functions of $N_B$ and $|a_{BF}|$.
Left panel: $u= 0.11$, $v= -0.0003$. Right panel: $u= 0.11$, $N_B=
5 \cdot 10^3$.}
\end{figure}

The coefficients $\alpha_1$ and $\alpha_2$ are plotted in Fig.
\ref{Fig3} as functions of the number of bosons $N_B$ and
scattering length $|a_{BF}|$. The chemical potential $\mu_B$, and
the number of bosons $N_B$ are fixed by $N_B= \int \rho_0({\bf r})
d^3r$. For small values of the $|a_{BF}|$ and $N_B$ the
coefficients $\alpha_1$ and $\alpha_2$ approach to one, while for
$N_B$ and $|a_{BF}|$ large enough, one observes, as expected,
important deviations due to the boson- fermion interaction
effects. The figure shows that the system will be in the collapse
regime both at sufficiently large $|a_{BF}|$ and $N_B$.

\begin{figure}
\includegraphics[width=8.8cm]{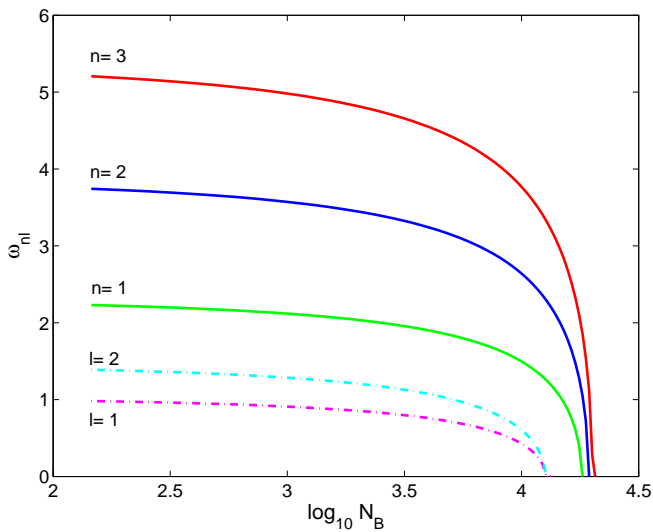}
\caption{\label{Fig4} Frequencies $\omega_{nl}$ (in units
$\omega_B$) of low- lying monopole modes $l= 0$, $n= 1,2,3$ (solid
line), dipole and quadrupole modes $n= 0$, $l=1, 2$ (dashed-dotted
line) as functions of the number of boson atoms $N_B$.}
\end{figure}

\begin{figure}
\includegraphics[width=8.8cm]{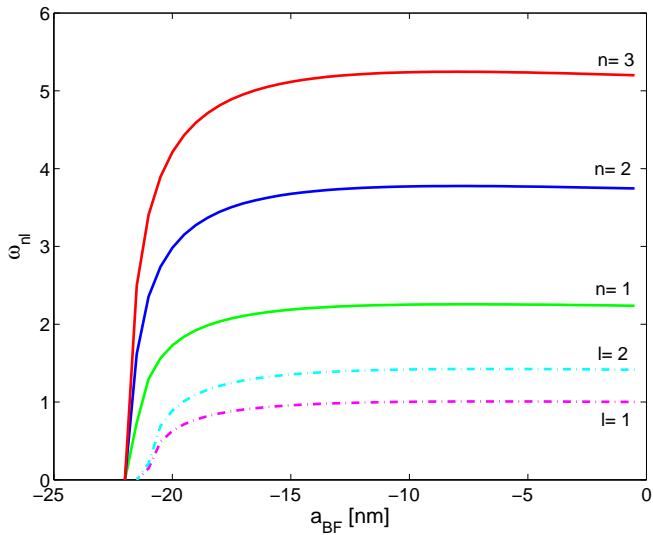}
\caption{\label{Fig5} Frequencies $\omega_{nl}$ (in units
$\omega_B$) of low-lying monopole modes $l= 0$, $n= 1,2,3$ (solid
line), dipole and quadrupole modes ($n= 0$, $l= 1, 2$)
(dashed-dotted line) as functions of the scattering length
$a_{BF}$ for $N_B= 5 \cdot 10^3$.}
\end{figure}

We will further limit the discussion to the case of excitations of
low multipolarity corresponding to the energy range $\omega_{nl}
\lesssim \mu$, where prediction \eqref{eigTF} is expected to be
accurate. To illustrate the dependence \eqref{eigTF} we consider
the low-lying monopole modes ($l= 0$) and the surface excitation
modes ($n= 0$), and keep the number of fermions fixed.

In Fig. \ref{Fig4} we show the evolution of $\omega_{nl}$ as a
function of number of bosons for $a_{BF}=-21.7$ nm. Three low-
lying monopole modes corresponding $n= 1,2,3$ and dipole and
quadrupole modes are shown. At increasing values of the boson
number the frequencies of modes display a monotonic decrease and
ultimately vanish in the vicinity of the collapse. Points of a
collapse are slightly different for different modes. This is a
result of a rather crude approximation that we chose for the
equilibrium density $\rho_0({\bf r})$. The second source of errors
comes from the TF approximation. Results based on the TF
approximation are in fact not adequate in the vicinity of $r
\simeq R_B$ and beyond. If one works with the full hydrodynamic
theory keeping the kinetic energy contributions in
\eqref{eig_inh}, one finds that $\rho_0({\bf r})$ does not
abruptly vanish at $r= R_B$ but exhibits a small tail which slowly
goes to zero. This tail is neglected in our approximation.
Nevertheless the behavior correctly display the situation which
occurs near the collapse. The similar frequency behavior of the
bosonic modes as a signature of the incipient collapse has been
obtained Capuzzi {\it et al.} with numerical calculations on the
basis of generalized hydrodynamics equations \cite{Capuzzi04}.

Fig. \ref{Fig5} shows the frequencies $\omega$ of the same low-
lying monopole and surface modes as in Fig. \ref{Fig4} as
functions of the scattering length $a_{BF}$ at fixed number of
bosons $N_B= 5 \cdot 10^3$. At larger values of $|a_{BF}|$ the
eigenfrequencies show softening as collapse is approached. In
contrast with the previous picture the points of the collapse are
approximately the same for different modes. At increasingly large
boson- fermion attraction the densities of the two species
increase in their overlap region and collapse occurs when this
attraction overcomes the Fermi kinetic pressure and the boson-
boson repulsion. Similar behavior has been revealed in
$^7$Li-$^6$Li mixture with numerical calculation of
eigenfrequencies \cite{Capuzzi04}.

Note that both components of the $^{87}$Rb-$^{40}$K mixture are
inside the same magnetic trap, but their trapping frequencies
differ considerably as a consequence of the large difference in
atomic masses. At this conditions the generalized  Kohn theorem
\cite{Dobson94} is satisfied when the species are uncoupled
($a_{BF} \to 0$) and each component enable to oscillate with its
own frequency. From Fig.~ \ref{Fig5} one can see that the dipole
mode frequency $\omega_D$ ($n= 0, l= 1$) in broad region on
$a_{BF}$ except near the collapse regime is found to be not
affected by the interactions and equals to the confining potential
frequency $\omega_B$. In fact this mode corresponds to the
oscillation of the center of mass of the Bose- gas with $\omega_D=
\omega_B$ driven by the external harmonic potentials. The same one
can say on the monopole $\omega_M$ and the quadrupole $\omega_Q$
mode frequencies. In this limit ($a_{BF} \to 0$) one recovers the
results $\omega_M= \sqrt{5} \omega_B$ ($n= 1, l= 0$) and
$\omega_Q= \sqrt{2} \omega_B$ ($n= 0, l= 2$) predicted by the
hydrodynamic theory of superfluid \cite{Stringari96}.

In Ref. \cite{Miyakawa2000, Sogo02} the collective oscillations
were calculated by the sum-rule approach and collisionless
random-phase approximation (RPA). Mixing angle of bosonic and
fermionic multipole operators was introduced so that the mixing
characters of the low- lying collective modes were studied as
functions of the boson- fermion interaction strength. For an
attractive boson- fermion interaction it was shown that the
low-lying monopole mode becomes a coherent oscillation of bosons
and fermions and shows a rapid decrease in the excitation energy
towards the instability point of the ground state. In contrast,
the quadrupole mode give indications of frequency increasing and
it contradicts the behavior of modes in Ref. \cite{Capuzzi04}. In
Ref. \cite{Capuzzi04} instead it was found that, as the mixture is
driven  towards collapse instability, the frequencies of the modes
of bosonic origin show a softening, which becomes most pronounced
in the very proximity of collapse. Explicit illustrations of these
trends were given for the monopolar spectra.

A comparison of the hydrodynamic spectra \cite{Capuzzi04} with the
spectra calculated on the basis of the effective Hamiltonian
suggests that eigenvalue equations with the dynamical coupling
between the two components yield the similar bosonic mode behavior
both as functions of $N_B$ and $a_{BF}$ as the hydrodynamic- type
equations \eqref{eig_inh} with the fermion degree of freedom being
integrated out. Softening of the bosonic mode spectrum is a
general signature of the dynamics of the Bose- Fermi mixture with
attraction between components arising from peculiar properties of
the ground state and instability point near the collapse
transition.

\section{Summary}

We have studied the collective mode frequencies in the boson-
fermion mixture. To this purpose we have used the effective
Hamiltonian for the Bose system, where the fermion degrees of
freedom are integrated out \cite{ChRyz04,cr5}. The effective
Hamiltonian incorporates the three- particle elastic collisions
induced by the boson- fermion interaction. In terms of coupled
eigenvalue density fluctuation equations for the bosonic and
fermionic components it means that we exclude fermion fluctuations
from the bosonic density fluctuation component resulting in more
involved structure of bosonic fluctuations.

This approach enables to account for interparticle interaction
strength in collective excitation frequencies and can be
considered as an extension of Stringari's solution
\cite{Stringari96} for Bose - Fermi mixtures. The analysis of the
mode frequencies as function of the number of boson atoms and the
boson- fermion coupling strength yields a dynamical condition for
a collapse of the system. The behavior of mode frequencies is
consistent with numerical analysis on the basis of generalized
hydrodynamics equations \cite{Capuzzi04}. The effective
Hamiltonian approach proves to be efficient tool for the analysis
of both the ground state and dynamical properties of boson-
fermion mixtures and can be considered as alternatives to
Capuzzi's {et al.} \cite{Capuzzi04} approach  for analysis boson
eigenfrequencies. We also note that the present study can be
extended to take into account finite temperature effect
\cite{Ryzhov08} on the excitations.

\section{Acknowledgement}
The work was supported by CRDF (A.M.B.) [Grant No. BF4M11] and the
Russian Foundation for Basic Research (V.N.R)(Grants No
08-02-00781 and No 10-02-00700), and Russian Federal Program
02.740.11.5160.


\begin{thebibliography}{99}

\bibitem{[1]}  M.\,N. Anderson, J.\,R. Ensher, M.\,R. Matthews
{\it et al.}, Science {\bf 269}, 198 (1995).
\bibitem{[2]} K.\,B. Davis, M.-O. Mewes, M.\,R. Andrews {\it et al.},
Phys. Rev. Lett. {\bf 75}, 3969 (1995).
\bibitem{[3]} C.\,C. Bradley, C.\,A. Sackett, and R.\,G. Hulet, Phys.
Rev. Lett. {\bf 78}, 985 (1997).

\bibitem{Bloch08} I. Bloch, J. Dalibard, and W. Zwerger,
Rev. Mod. Phys. {\bf 80}, 885 (2008).

\bibitem{Giorgini08} S. Giorgini, L.\,P. Pitaevskii,
and S. Stringari, Rev. Mod. Phys. {\bf 80}, 1215 (2008).

\bibitem{cr1} S.\,T. Chui, V.\,N. Ryzhov, and E.\,E. Tareyeva, JETP
{\bf 91},1183(2000).

\bibitem{cr2} S.\,T. Chui, V.\,N. Ryzhov, and E.\,E. Tareyeva,
Phys. Rev. A {\bf 63}, 023605  (2001).

\bibitem{cr3} S.\,T. Chui, V.\,N. Ryzhov, and E.\,E. Tareyeva, J.
Phys.: Condensed Matter {\bf 14}, L77 (2002).

\bibitem{cr4} S.\,T. Chui, V.\,N. Ryzhov, and E.\,E. Tareyeva, Jetp
Letters, {\bf 75}, 233 (2002).

\bibitem{sci99} B. DeMarco and D.\,S. Jin, Science {\bf 285}, 1703
(1999).
\bibitem{sci01} A.\,G. Truscott, K.\,E. Strecker, W.\,I. McAlexander
{\it et al}, Science {\bf 291}, 2570 (2001).

\bibitem{prl1} F. Schreck, L. Khaykovich, K.\,L. Corwin {\it et
al}, Phys. Rev. Lett. {\bf 87}, 080403 (2001).
\bibitem{NaLi} Z. Hadzibabic, C.\,A. Stan, K. Dieckmann {\it et
al}, Phys. Rev. Lett. {\bf 88}, 160401 (2002).
\bibitem{Modugno02} G. Modungo, G. Roati, F. Riboli {\it et al},
Science {\bf 297}, 2240 (2002).
\bibitem{KRb} J. Goldwin, S.\,B. Papp, B. DeMarco, and D.\,S. Jin,
Phys. Rev. A {\bf 65}, 021402(R) (2002).

\bibitem{Nygaard99} N. Nygaard and K. Molmer, Phys. Rev. A
\textbf{59}, 2974 (1999).

\bibitem{Yi01} X.X. Yi and C.P. Sun, Phys. Rev. A
\textbf{64}, 043608 (2001).

\bibitem{Minguzzi2000} A. Minguzzi and M.P. Tosi, Phys. Lett. A {\bf 268}, 142 (2000).
\bibitem{Akdeniz02} Z. Akdeniz, A. Minguzzi, P. Vignolo and M.P. Tosi, Phys. Rev. A {\bf 66}, 013620 (2002).



\bibitem{Capuzzi02} P. Capuzzi and E.S. Hernandez, Phys. Rev. A  \textbf{66}, 035602 (2002);

\bibitem{Roth02} R. Roth and H. Feldmeier, Phys. Rev. A {\bf 65},
021603(R) (2002); R. Roth, {\it ibid.} {\bf 66}, 013614 (2002).

\bibitem{Molmer98} K. Molmer, Phys. Rev. Lett. \textbf{80}, 1804 (1998).

\bibitem{jezek} D.M. Jezek, M. Barranco, M. Guilleumas, R. Mayol,
and M. Pi, Phys. Rev. A {\bf 70}, 043630 (2004).

\bibitem{Stoof2000} M.J. Bijlsma, B.A. Heringa, and H.T.C. Stoof, Phys. Rev. A {\bf 61},
053601 (2000).


\bibitem{Ospelkaus06} C. Ospelkaus, S. Ospelkaus, K. Sengstock, and K. Bongs, Phys. Rev. Lett. {\bf 96}, 020401 (2006).

\bibitem{Capuzzi01} P. Capuzzi and E.S. Hernandez, Phys. Rev. A
\textbf{64}, 043607 (2001); J. Low Temp. Phys. \textbf{126}, 425
(2002).
\bibitem{Maruyama05} T. Maruyama, H. Yabu, and T. Suzuki, Phys. Rev. A \textbf{72},
013609 (2005).


\bibitem{Capuzzi03} P. Capuzzi, A. Minguzzi, and M.P. Tosi, Phys. Rev. A \textbf{67},
053605 (2003).
\bibitem{Capuzzi03_2} P. Capuzzi, A. Minguzzi, and M.P. Tosi, Phys. Rev. A \textbf{68},
033605 (2003).

\bibitem{Capuzzi04} P. Capuzzi, A. Minguzzi, and M.P. Tosi, Phys. Rev. A \textbf{69}, 053615 (2004).


\bibitem{Miyakawa2000} T. Miyakawa, T. Suzuki, and H. Yabu, Phys. Rev. A \textbf{62},
063613 (2000).

\bibitem{Sogo02} T. Sogo, T. Miyakawa, T. Suzuki, and H. Yabu, Phys. Rev. A \textbf{66},
013618 (2002).




\bibitem{Stringari96} S. Stringari, Phys. Rev. Lett. {\bf 77}, 2360 (1996).

\bibitem{Edwards96} M. Edwards, P.\,A. Ruprecht, K. Burnett, R.\,J. Dodd and C.\,W. Clark,
Phys. Rev. Lett. {\bf 77}, 1671 (1996).

\bibitem{Ueda98} M. Ueda and A. Leggett, Phys. Rev. Lett. {\bf 80}, 1576 (1998).


\bibitem{ChRyz04} S.\,T. Chui and V.\,N. Ryzhov, Phys. Rev. A {\bf
69}, 043607 (2004).

\bibitem{cr5} S.\,T. Chui, V.\,N. Ryzhov, and E.\,E. Tareyeva, JETP Lett.
{\bf 80}, 274 (2004).

\bibitem{popov1} V.\,N. Popov, {\it Functional Integrals in Quantum Field Theory
and Statistical Physics} (Reidel, Dordrecht, 1983).

\bibitem{stoof1} H.\,T.\,C. Stoof, in {\it Proceedings of the Les Houches Summer School
on Coherent Atomic Matter Waves, Session LXXII, 1999}, Ed. by R.
Kaiser, C. Westbrook, and F. David (Springer, Berlin, 2001), pp.
219-316; e-print arXiv: cond-matt/9910441.


\bibitem{Dalfovo99} F. Dalfovo, S. Giorgini, L.\,P. Pitaevskii,
and S. Stringari, Rev. Mod. Phys. {\bf 71}, 463 (1999).

\bibitem{Kagan9698} Yu. Kagan, G.\,V. Shlyapnikov, and J.\,T.\,M.
Walraven, Phys. Rev. Lett. {\bf 76}, 2670 (1996); Yu. Kagan,
A.\,E. Muryshev, and G.\,V. Shlyapnikov, {\it ibid.} {\bf 81}, 933
(1998).

\bibitem{Fetter96} A.L. Fetter, Phys. Rev. A \textbf{53}, 4245 (1996).

\bibitem{Butts97} D.\,A. Butts and D.\,S. Rokhsar, Phys. Rev. A {\bf 55}, 4346 (1997).

\bibitem{Ryzhov07} A.M. Belemuk, V.N. Ryzhov, and S.-T. Chui, Phys. Rev. A \textbf{76},
013609 (2007).

\bibitem{cr6} A.\,M. Belemuk, V.\,N. Ryzhov, and S.-T. Chui, JETP Lett. {\bf 84}, 294 (2006).

\bibitem{PRA06} A.M. Belemuk, N.M. Chtchelkatchev, V.N. Ryzhov, and S.-T. Chui, Phys. Rev. A {\bf
73}, 053608 (2006).


\bibitem{Brachet99} C. Huepe, S. Metens, G. Dewel, P. Borckmans, and M.E. Brachet,
Phys. Rev. Lett. {\bf 82}, 1616 (1999).

\bibitem{Stoof96} M. Houbiers and H.T.C. Stoof, Phys. Rev. A {\bf 54},
5055 (1996).

\bibitem{Dobson94} J.F. Dobson, Phys. Rev. Lett. {\bf 73}, 2244 (1994).

\bibitem{Ryzhov08} A.M. Belemuk and V.N. Ryzhov, JETP Lett. 87,
376 (2008)

\bibitem{Lifshitz80} E.\,M. Lifshitz and L.\,P. Pitaevskii, {\it Statistical Physics, Part 2}
(Pergamon Press, Oxford, 1980).

\bibitem{Abramowitz} {\it Handbook of Mathematical Functions with Formulas, Graphs, and Mathematical Tables},
edited by M. Abramowitz and I.A. Stegun, Natl. Bur. Stand. Appl.
Math. Ser. No. 55 (U.S. GPO, Washington, DC, 1968), Chaps. 15 and
22.

\end{thebibliography}
\end{document}